\documentclass[a4paper,10pt,twoside,slidetop,cjk]{cpc-hepnp}
\usepackage{CJK,upgreek,fancyhdr}
\usepackage{multicol}
\usepackage{graphicx}
\usepackage{booktabs}
\usepackage{amssymb,bm,mathrsfs,bbm,amscd}
\usepackage[tbtags]{amsmath}
\usepackage{lastpage}
\usepackage{color}
\usepackage{leftidx}

\begin{document}
\begin{CJK*}{GB}{gbsn}

\fancyhead[c]{\small Submitted to Chinese Physics C}
\fancyfoot[C]{\small 010201-\thepage}

\footnotetext[0]{Received 2 June 2017}

\title{Estimating proton radius and proportion of other non-perturbative components in the proton by the Maximum Entropy Method
\thanks{This work is supported by the National Basic Research Program of China (973 Program)
2014CB845406. } }

\author{%
     Chengdong Han(º«³É¶°)$^{1,2}$%
\quad Xurong Chen(³ÂÐñÈÙ)$^{1)}$\email{xchen@impcas.ac.cn}%
}
\maketitle

\address{%
$^1$ Institute of Modern Physics, Chinese Academy of Sciences, Lanzhou 730000, China\\
$^2$ University of Chinese Academy of Sciences, Beijing 100049, China\\
}

\begin{abstract}
In this paper, we apply the Maximum Entropy Method to estimate the proton radius and determine the valence quark distributions in the proton at extremely low resolution scale Q$^{2}_{0}$. Using the simplest functional form of the valence quark distribution and standard deviations of quark distribution functions in the estimation of the proton radius, we obtain a quadratic polynomial for the relationship between the proton radius and the momentum fraction of other non-perturbative components in the proton. The proton radii are approximately equal to the muonic hydrogen experimental result $r_p$ = 0.841~fm and the CODATA analysis $r_p$ = 0.877 fm when the other non-perturbative components account for 17.5\% and 22.3\% respectively. We propose $``ghost~~matter"$ to explain the difference in other non-perturbative components (4.8\%) that the electron can detect.
\end{abstract}



\footnotetext[0]{\hspace*{-3mm}\raisebox{0.3ex}{$\scriptstyle\copyright$}2017
Chinese Physical Society and the Institute of High Energy Physics
of the Chinese Academy of Sciences and the Institute
of Modern Physics of the Chinese Academy of Sciences and IOP Publishing Ltd}%

\begin{multicols}{2}

\section{Introduction}
The proton radius puzzle \cite{1,2,3} is an open problem in both particle physics and nuclear physics. 
Due to the emergence of precision Lamb shift measurements of the muonic hydrogen atom ($\mu$p), it has received much attention in the last few years. The muonic hydrogen result for the proton radius is $r_p$ = 0.84087 $\pm$ 0.00039 fm \cite{4,5}, which differs by about 7.9 standard deviations from the CODATA value of $r_p$ = 0.8775 $\pm$ 0.0051 fm \cite{6}. The CODATA value comes from the analysis of electron-proton scattering data and electronic hydrogen spectroscopy. This disagreement, called the proton radius puzzle, has not been solved yet.
There are many proposals to resolve the proton radius puzzle via novel interactions for ep and $\mu$p \cite{7,8} or by applying some algorithms, such as polynomial fits~\cite{9}, Bayesian inference~\cite{10}, etc.

In this paper, we use the Maximum Entropy Method (MEM) to extract the proton radius. This approach is inspired by Ref.~\cite{11}. We consider the entropy as a function of the proton radius.  When considering other non-perturbative components for the proton radius correction, we get reasonable results for the proton radius and the proportion of these other non-perturbative components in the proton. The other non-perturbative components include the cloud sea in the $\pi$ cloud model \cite{12,13,14,15}, the five-quark components of the proton \cite{16,17,18,19}, the intrinsic sea quarks in light front QCD theory \cite{20,21,22}, and the connected sea quarks in LQCD \cite{23,24,25}.
Finally, we compare the predicted up and down valence quark momentum distributions with the global QCD fits CT10 \cite{26}, MSTW08(LO) \cite{27} and IMParton16 \cite{28}, and find that our obtained up and down valence quark momentum distributions are nearly consistent with the popular parton distribution functions from QCD global analyses.

\section{A naive non-perturbative input from the quark model}

The quark model has been a great success in the study of hadron spectra and reaction dynamics in the last several decades, and it has revealed the internal symmetry of hadrons. The proton consists of three colored quarks at a low scale $Q_{0}^{2}$. Thus, a naive nonperturbative description of the proton includes only these three valence quarks \cite{29,30,31,32}, which is the simplest initial parton distribution.
The simplest functional form used to parametrize the valence quark distribution is the time-honored canonical parametrization $f(x)=Ax^{B}(1-x)^{C}$ \cite{33}, which provides a good approximation to the valence distribution at large $x$ (the Bjorken scaling variable). Therefore we use the naive inputs
\begin{equation}
\begin{aligned}
&u_v(x,Q_0^2)=A_u x^{B_u}(1-x)^{C_u},\\
&d_v(x,Q_0^2)=A_d x^{B_d}(1-x)^{C_d}.
\end{aligned}
\label{Parametrization}
\end{equation}
In the proton there are two $u$ quarks and one $d$ quark. This yields the valence sum rules for the naive non-perturbtive inputs
 \begin{equation}
\int_0^1 u_v(x,Q_0^2)dx=2,
\int_0^1 d_v(x,Q_0^2)dx=1.
\label{ValenceSum}
\end{equation}

When considering the presence of other non-perturbative components in the proton \cite{34}, we define $\Delta~$($0\leq\Delta<1$) as the contribution of other non-perturbative proton constituents to the momentum. Then the momentum sum for valence quarks is $1-\Delta$. The momentum sum rule for valence quarks using the naive input becomes
\begin{equation}
\int_0^1 x[u_v(x,Q_0^2)+d_v(x,Q_0^2)]dx=1-\Delta.
\label{MomentumSum}
\end{equation}
The valence quarks carry all the momentum of the proton when $\Delta=0$. Hence, the momentum sum rule for valence quarks using the naive input is
\begin{equation}
\int_0^1 x[u_v(x,Q_0^2)+d_v(x,Q_0^2)]dx=1.
\label{MomentumSum}
\end{equation}

\section{Standard Deviations of Quark Distribution Functions and the Maximum Entropy Method}
\label{SecIII}
The valence quarks in a proton are confined to a small spatial region. According to Heisenberg's uncertainty principle, the momenta of the quarks in the proton are therefore uncertain, and are described by probability density distributions. The Heisenberg uncertainty principle is
\begin{equation}
\sigma_X\sigma_P \ge \frac{\hbar}{2}.
\label{Uncertainty}
\end{equation}
The uncertainty relation is
$\sigma_X\sigma_P = \hbar/2$ for
a quantum harmonic oscillator in the ground state in quantum mechanics.
In order to constrain the standard deviation of the quark momentum distributions,
$\sigma_X\sigma_P = \hbar/2$ is assumed for the three initial valence quarks
in our analysis instead of $\sigma_X\sigma_P \ge \hbar/2$.
$\sigma_X$ is related to the radius of the proton.
A simple estimation of $\sigma_X$ is $\sigma_X = (2\pi R^3/3)/(\pi R^2)=2R/3$,
where R is a proton radius.
$\sigma_X$ of each $u$ valence quark is divided by
$2^{1/3}$ as there are two $u$ valence quarks in the spatial region.
Then we have $\sigma_{X_d}=2R/3$ and $\sigma_{X_u}=2R/(3\times 2^{1/3})$ \cite{11}.

The Bjorken scaling variable $x$ is the momentum fraction carried by one parton in the proton. In the proton rest frame, we use the energy ratio instead. Therefore, in this paper we assume that the standard deviation
of x at an extremely low resolution scale $Q^{2}_{0}$ is written as
\begin{equation}
\sigma_x = \frac{\sigma_P}{P_p},
\label{xDeviation}
\end{equation}
where $P_p$ is the energy of the proton, which equals the mass (0.938 GeV \cite{35}) of the proton. Constraints for the valence quark distributions from QCD confinement and the Heisenberg uncertainty principle are expressed in Ref.~\cite{11}.
Then by applying the MEM, we can estimate the proton radius and get the valence quark distributions from the constraint equations discussed above. The generalized information entropy of the valence quarks is defined as
\begin{equation}
\begin{aligned}
S=&-\int_0^1 [ 2 \frac{u_v(x,Q_0^2)}{2}ln(\frac{u_v(x,Q_0^2)}{2}) \\
  &+ d_v(x,Q_0^2)ln(d_v(x,Q_0^2)) ] dx.
\end{aligned}
\label{EntropyFor}
\end{equation}

In this paper we will consider two cases corresponding to distinct non-perturbative inputs. In the first case, the proton does not contain the other non-perturbative components, so the valence quarks carry all the momentum of the proton. Then the momentum sum equals 1 for valence quarks at $Q_0^2$.  The other case is that there are also flavor-asymmetric sea components \cite{34} in the naive non-perturbative input. Namely, the flavor-asymmetric sea components carry part of the momentum of the proton.  By considering the above two cases, we are able to estimate the values of the proton radius and the distribution functions of the valence quarks.  Then we find an expression that relates the proton radius to the proportion of other non-perturbative components in the proton.  Finally, we compare our predicted valence quark momentum distributions with the latest global fits of parton distribution functions.

\section{Results and Discussion}
\begin{table*}
\begin{center}
\tabcaption{ \label{tab2}  $\Delta$, $B_{d}$, and $R$ are the momentum fraction of other non-perturbative components in the proton, the parameter of the $u$ quark distribution function (1), and the value of the proton radius, respectively.}
\footnotesize
\begin{tabular*}{170mm}{@{\extracolsep{\fill}}ccccccccccccccccccccccccccccccccccccccccccccccccccccccccccccccc}
\hline\hline
$~\Delta~$      &$0$           &$0.02$               &$0.04$              &$0.06$               &$0.08$
                               &$0.1$                &$0.12$ \\\hline

$~B_{d}~$  &$3.411\times10^{-2}$&$2.529\times10^{-2}$&$1.947\times10^{-2}$ &$1.371\times10^{-2}$ &$9.25\times10^{-3}$
                               &$2.09\times10^{-3}$  &$-2.15\times10^{-3}$            \\

$~R~$      &$0.745$            &$0.754$             &$0.763$             &$0.773$             &$0.783$
                               &$0.794$             &$0.805$ \\

\hline\hline
$~\Delta~$                &$0.14$              &$0.15$               &$0.16$
                          &$0.17$              &$0.175$              &$0.18$                &$0.19$                 \\\hline

$~B_{d}~$        &$-6.08\times10^{-3}$         &$-8.60\times10^{-3}$ &$-8.28\times10^{-3}$
                         &$-7.84\times10^{-3}$ &$-9.03\times10^{-3}$ &$-1.032\times10^{-2}$ &$-1.173\times10^{-2}$      \\

$~R~$                   &$0.817$               &$0.824$              &$0.830$

                        &$0.837$               &$0.841$             &$0.845$                &$0.852$                  \\
\hline\hline
$~\Delta~$               &$0.20$               &$0.21$               &$0.22$              &$0.223$               &$0.23$
                         &$0.24$               &$0.25$                                     \\\hline

$~B_{d}~$       &$-1.309\times10^{-2}$&$-1.404\times10^{-2}$&$-1.337\times10^{-2}$&$-1.483\times10^{-2}$ &$-1.526\times10^{-2}$        &$-1.588\times10^{-2}$                               &$-1.621\times10^{-2}$                                     \\

$~R~$                  &$0.859$             &$0.867$             &$0.875$            &$0.877$             &$0.883$
                       &$0.892$             &$0.901$                                     \\
\hline \hline
\end{tabular*}%
\end{center}
\end{table*}

We consider the information entropy S($B_d$, R) as a function of the variable $B_d$ and the proton radius R. The momentum sum for valence quarks equals 1 in Eq.~(4).
\begin{center}
\includegraphics[width=9cm]{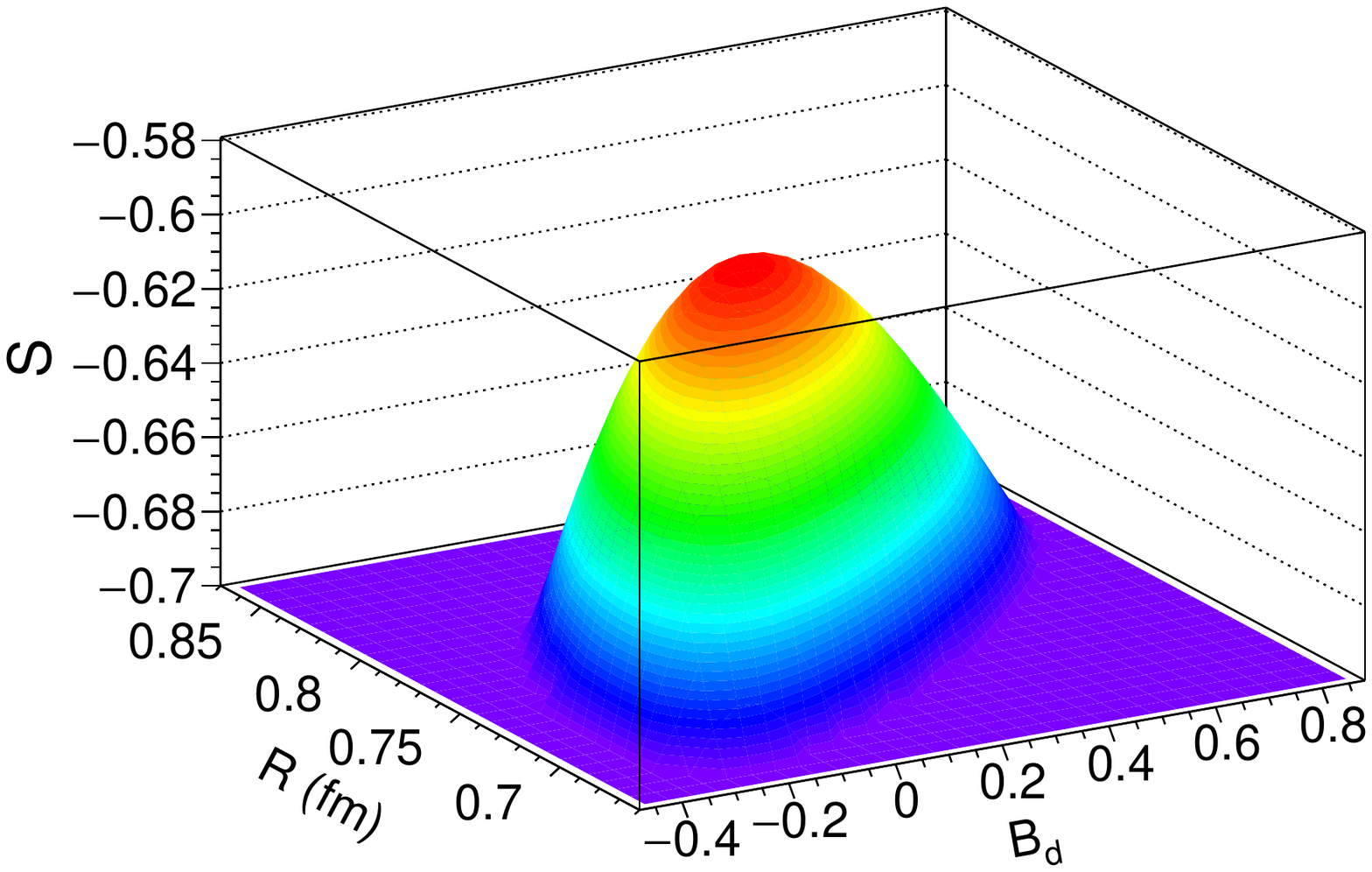}
\figcaption{\label{fig1} The information entropy $S$ plotted as a function of the parameters $B_d$ and R ($\Delta$ = 0). }
\end{center}
With the constraints given in Eqs. (2) , (4) , (5) , and (6), there are two free parameters ($B_d$ and R) left for the parametrized naive nonperturbative input. Figure 1 shows the information entropy of the valence quark distributions of the proton at the starting scale as a function of the parameters R and $B_d$. We assume that valence quarks carry all of the momentum of the proton in Fig. 1. Using the MEM, we find that the R is optimized at 0.745 fm, and $B_d$ is 0.03411. The corresponding valence quark distributions are as follows
\begin{equation}
\begin{aligned}
&u_v(x,Q_0^2)=2.13911 x^{-0.21026}(1-x)^{0.43090},\\
&d_v(x,Q_0^2)=2.69719 x^{0.03411}(1-x)^{1.54750}.
\end{aligned}
\label{Parametrization}
\end{equation}

Analysing Fig. 1, we find that the proton radius is smaller than the muonic hydrogen result and the CODATA analysis. Next, flavor-asymmetric sea components \cite{34} will be included and found to be a rather better non-perturbative input. Namely, we will allow the proton to contain other non-perturbative components in addition to the three valence quarks. In order to better explain the value of the proton radius estimated by the MEM, we should take into account the contribution of other non-perturbative components to the proton radius. The non-perturbative components include the cloud sea in the pion cloud \cite{12,13,14,15}, the five-quark components in the proton \cite{16,17,18,19}, the intrinsic sea quarks in light front QCD theory \cite{20,21,22}, and the connected sea quarks in LQCD \cite{23,24,25}.
These non-perturbative components lead to the flavor-asymmetric sea quarks in the proton.
Y.-J. Zhang et al.~\cite{36,37} proved the light flavor sea quark asymmetry of the proton, i.e. $\bar{u}\neq\bar{d}$, by the principle of detailed balance and the balance model, which is based on a purely statistical effect. Furthermore, they calculated the strange content of the proton using the balance model under the equal probability assumption.
In addition, muon-hydrogen and muon-deuterium deep inelastic scattering measurements also find an excess of $\overline{d}$ and $\overline{u}$ sea quarks \cite{38}. On the other hand, the fact that the value of the proton radius is smaller than the muonic hydrogen result and the CODATA analysis when valence quarks carry all the momentum of proton supports the case for the pion cloud model or other models. So we need to consider the proportion of momentum carried by other non-perturbative components of the proton. Then we scan the momentum fraction ($\Delta$) for other non-perturbative components from 0.00 to 0.25.
\begin{center}
\includegraphics[width=7cm]{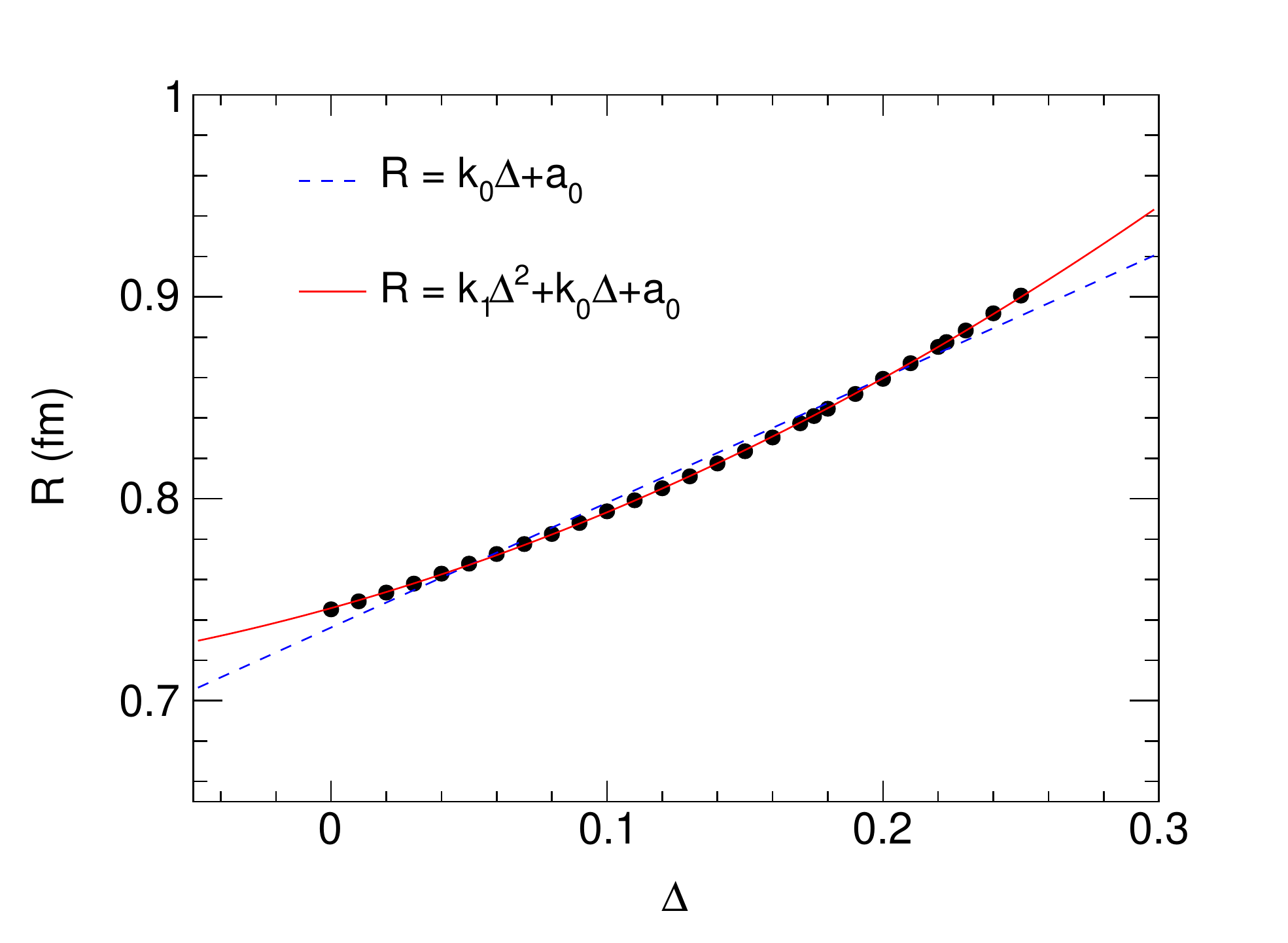}
\figcaption{\label{fig2}The relationship between the proton radius and the momentum fraction of other non-perturbative components in the proton.~The fitting curves are a linear function (dashed line) and a quadratic polynomial (solid line) respectively. }
\end{center}

\begin{center}
\tabcaption{ \label{tab2}  The fitted parameters with their errors.}
\footnotesize
\begin{tabular*}{80mm}{c@{\extracolsep{\fill}}cccccccccccccccccccccccccccccccccccc}
\hline\hline
$$             &$k_{1}$              &$k_{0}$               &$a_{0}$~                        \\\hline
$Fit~(9)~$    &$$ --                &$0.618\pm0.012$       &$0.736\pm0.012$~                 \\
$Fit~(10)~$   &$0.944\pm0.015$      &$0.380\pm0.004$       &$0.746\pm0.000$~                 \\
\hline\hline
\end{tabular*}
\vspace{0mm}
\end{center}

In Table 1 and Fig. 2, R represents the proton radius in fm and $\Delta$ represents the momentum fraction of other non-perturbative components in the proton.~The proton radius is approximately equal to the muonic hydrogen experimental result and the CODATA analysis when other non-perturbative components account for 17.5\% and 22.3\% respectively. Through the fitting of the data points, we get an expression for the relationship between the proton radius and the momentum fraction of other non-perturbative components in the proton
\begin{equation}
\begin{aligned}
&R=k_{0}\Delta+a_{0}.
\end{aligned}
\label{Parametrization}
\end{equation}
\begin{equation}
\begin{aligned}
&R=k_{1} \Delta^{2}+k_{0}\Delta+a_{0},
\end{aligned}
\label{Parametrization}
\end{equation}
where $k_{0}$, $k_{1}$ and $a_{0}$ are adjustable parameters. Equations (9) and (10) are the relationship between the proton radius and the momentum fraction of other non-perturbative components in the proton; they are a linear function and a quadratic function, respectively. Through fitting the data points in Table 1, where Table 2 gives the fitted parameters with their errors, it is apparent that Eq.~(10) has a better goodness-of-fit.
\begin{center}
\includegraphics[width=8cm]{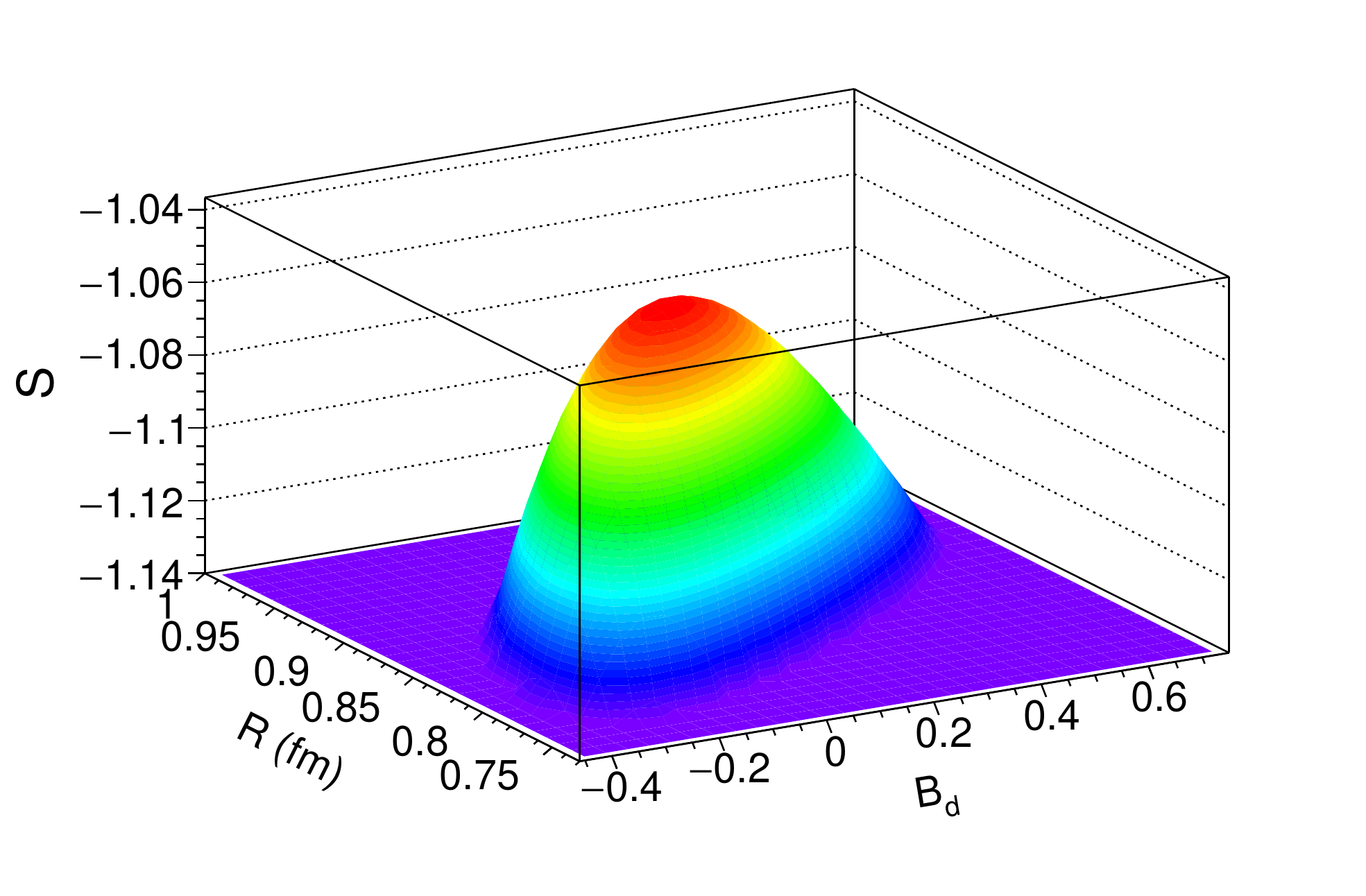}
\figcaption{\label{fig4} Information entropy $S$ plotted as a function of the parameters $B_d$ and R ($\Delta$ = 0.175). }
\end{center}

Figure 3 shows the information entropy of the valence quark distribution of the proton at the starting scale as a function of the parameters R and $B_d$. According to the MEM, R is optimized at 0.841 fm, and $B_d$ is 0.00903 when the other non-perturbative components occupy 17.5 percent of the proton's momentum. Then the corresponding valence quark distributions are as follows:
\begin{equation}
\begin{aligned}
&u_v(x,Q_0^2)=2.75223 x^{-0.20512}(1-x)^{0.91502},\\
&d_v(x,Q_0^2)=3.11320 x^{0.00903}(1-x)^{2.16667}.
\end{aligned}
\label{Parametrization}
\end{equation}

Next we consider the information entropy of the valence quark distribution of the proton at the starting scale as a function of the parameters R, $B_d$ and M. Here the momentum sum M in Eq.~(3) is a free parameter. By fitting, R is optimized at 0.745 fm, $B_d$ at 0.0341 and M is 1. The proton radius R, $B_d$ and M are equivalent to the situation in which we take the proton radius R and $B_d$ as free parameters and assume that valence quarks take the total momentum of the proton. So the corresponding valence quark distributions are equivalent to the distributions in Eq.~(8).

Through analysis of the above results, we find that the proton radius is smaller than muonic hydrogen result and the CODATA result if the proton only contains three valence quarks.~The value of proton radius which we estimated using the MEM is approximately equal to the measured value of the muonic hydrogen ($\mu$p) experiment and the CODATA analysis when the other non-perturbative components account for 17.5\% and 22.3\% of the proton momentum, respectively.
\begin{center}
\includegraphics[width=9cm]{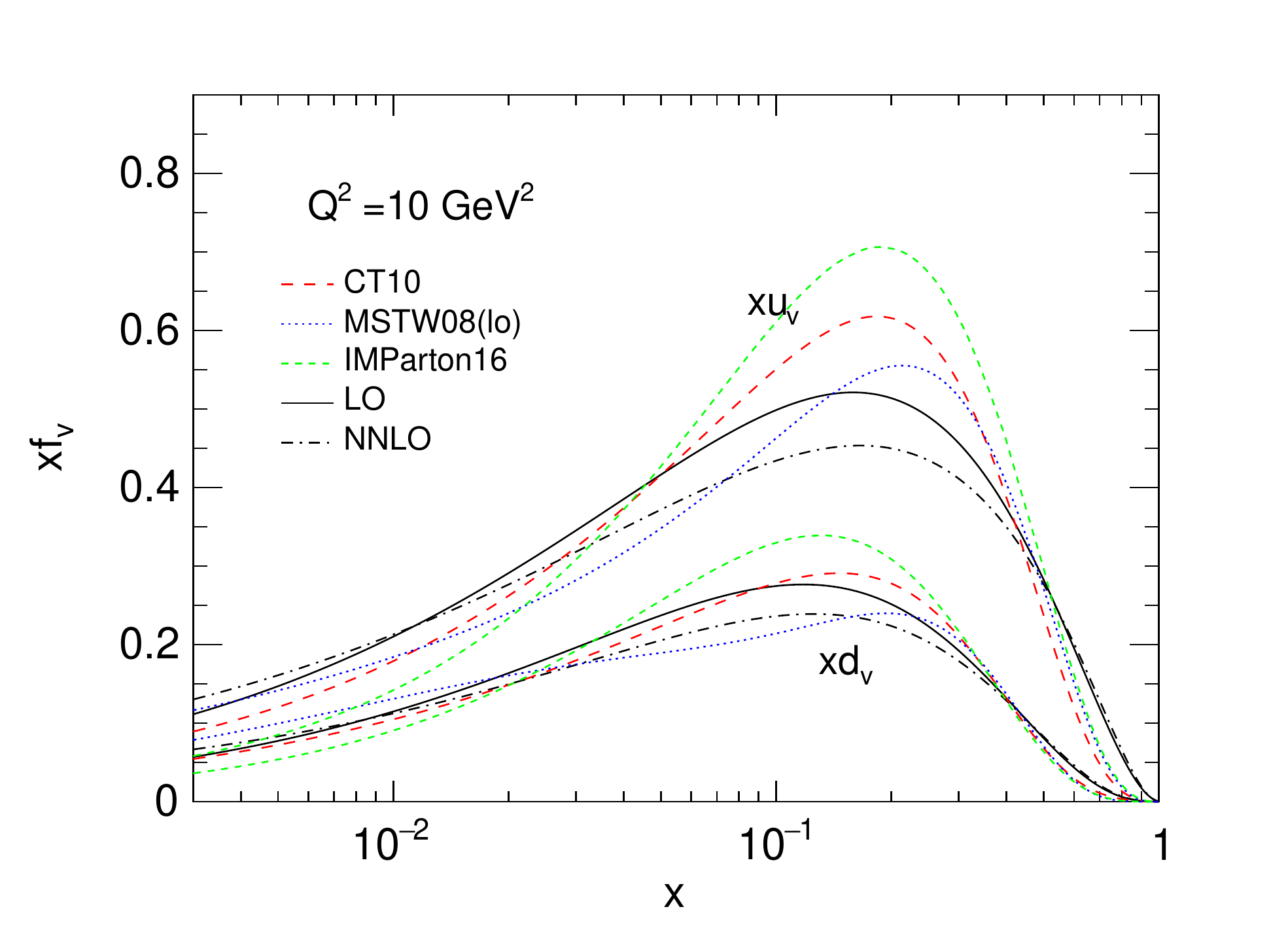}
\figcaption{\label{fig4} Comparisons of our predicted up and down valence quark momentum distributions from Eq.~(8) (solid and dot-dashed lines) with
global QCD fits CT10 (dashed lines) \cite{26} , MSTW08(LO) (dotted lines) \cite{27} and IMParton16 (short dashed lines) \cite{28}.}
\end{center}

\begin{center}
\includegraphics[width=9cm]{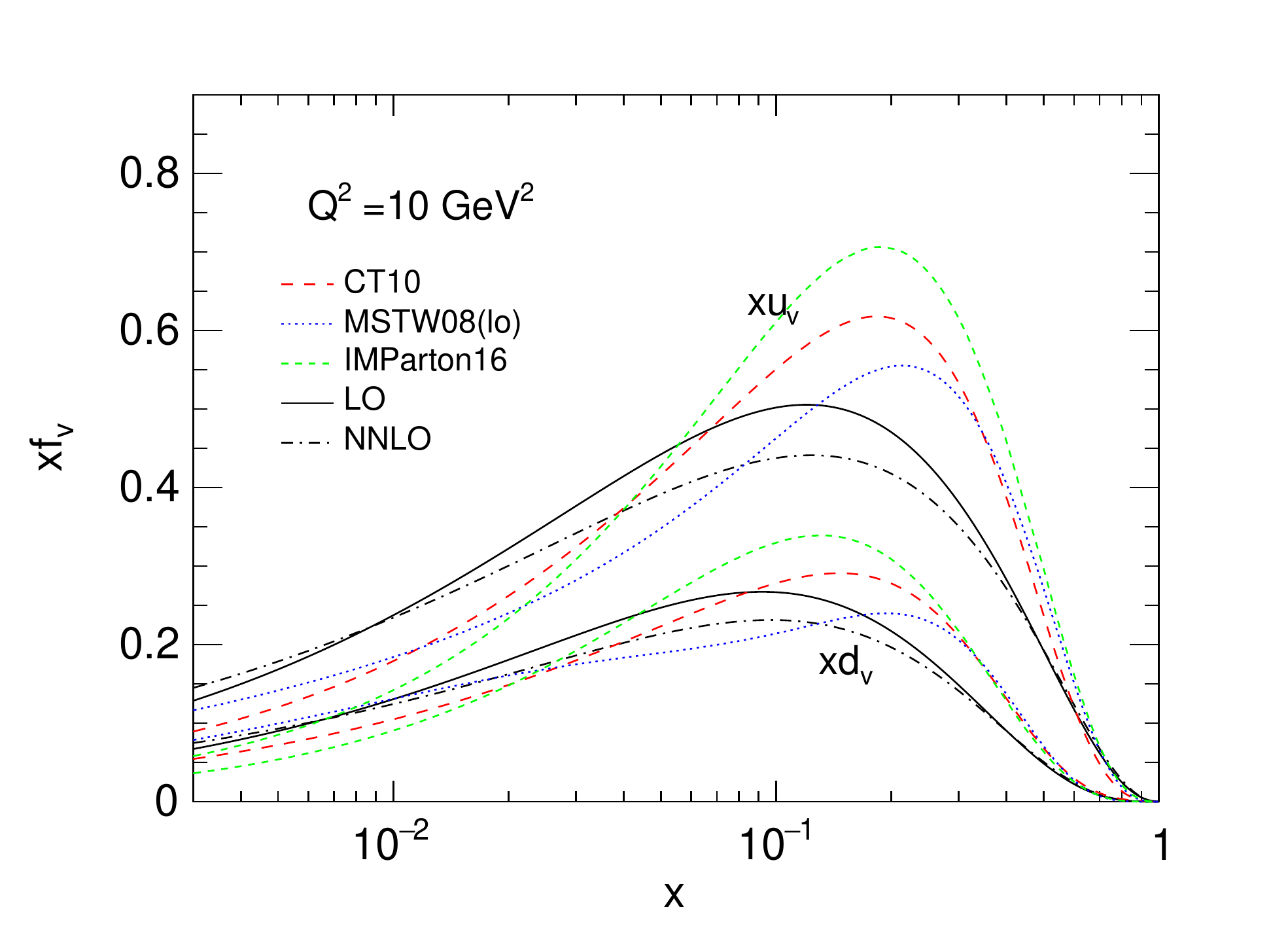}
\figcaption{\label{fig5} Comparisons of our predicted up and down valence quark momentum distributions from Eq.~(11) (solid and dot-dashed lines) with
global QCD fits CT10 (dashed lines) \cite{26} , MSTW08(LO) (dotted lines) \cite{27} and IMParton16 (short dashed lines)\cite{28}.}
\end{center}

Figures 4 and 5 show comparisons of the up and down valence quark momentum distributions we obtained, multiplied by $x$ (the Bjorken scaling variable). These predicted up and down valence quark momentum distributions are compared with the global fits from CT10 \cite{26} , MSTW08 \cite{27} and IMParton16 \cite{28} at $Q^{2}$ = 10 GeV$^{2}$. Figure 4 shows a comparison of our predicted up and down valence quark momentum distributions from Eq.~(8) with the global QCD fits CT10 , MSTW08(LO) and IMParton16. Figure 5 shows a comparison of our predicted up and down valence quark momentum distributions from Eq.~(11) with the global QCD fits CT10 , MSTW08(LsO) and IMParton16.  The up and down valence quark momentum distributions which we obtained are basically consistent with the popular parton distribution functions from QCD global analyses.

\section{Summary}

We have attempted to determine the proton radius and parton distribution functions using the MEM and to determine the presence of other non-perturbative components of the proton.
Previously, Y.-H. Zhang et al. \cite{39} calculated the proton radius and parton ditribution functions through a statistical approach, where the radius value is close to the result by Mac and Ugaz \cite{40} with the consideration of first-order QCD corrections.
 We found a quadratic polynomial that relates the proton radius and the proportion of other non-perturbative components in the proton. The obtained proton radius, which considers the pion cloud model or intrinsic sea quarks or connected sea quarks or other non-perturbative components, is consistent with the muonic hydrogen atom result and the CODATA analysis when the other non-perturbation components account for 17.5\% and 22.3\%, respectively. Moreover, the valence quark distributions are basically consistent with the experimental observations from high energy lepton probes and PDFs from global analyses. However, the MEM is applicable for obtaining the proton radius and valence quark distributions with the least bias in situations where detailed information is not given. Through calculation and analysis, we find that the proton radius, the proportion of other non-perturbative components in the proton and the valence quark distributions obtained by the MEM are reasonable.

The difference in the non-perturbative components between the muonic hydrogen atom result and the CODATA analysis is 4.8\%. An interesting question is whether an electron from the CODATA analysis is more likely to detect other non-perturbative components than the muon from the muonic hydrogen atom result.
We propose $``ghost~~matter"$ to explain the difference in other non perturbative components (4.8\%) that the electron can detect,
which shows that using muons and electrons as a probe will correspond to different proton radii.
Furthermore, the MEM can also be used to make theoretical predictions, estimate the radii of other hadrons, and get the proportion of other non-perturbative components inside.

\noindent{\bf Acknowledgments}:
The authors would like to thank Rong Wang for helpful and fruitful suggestions.
This work is supported by the National Basic Research Program of China (973 Program)
2014CB845406.
\end{multicols}

\vspace{-1mm}
\centerline{\rule{80mm}{0.1pt}}
\vspace{2mm}
\begin{multicols}{2}

\end{multicols}

\clearpage
\end{CJK*}

\begin{thebibliography}{90}

\vspace{3mm}

\bibitem{1}
F. Dahia, A. S. Lemos, Eur. Phys. J. C 76, 435 (2016).

\bibitem{2}
I.T. Lorenz, H.-W. Hammer, and Ulf-G. Mei${\ss}$ner, Eur. Phys. J. A 48: 151 (2012).

\bibitem{3}
Carl E. Carlson, arXiv:1502.05314v1.

\bibitem{4}
 R. Pohl, A. Antognini, F. Nez, F. D. Amaro, F. Biraben, et al. Nature. 466, 213 (2010).

\bibitem{5}
A. Antognini, F. Nez, K. Schuhmann, F. D. Amaro, F. Biraben, et al. Science. 339, 417 (2013).

\bibitem{6}
P. J. Mohr, B. N. Taylor, and D. B. Newell, Rev. Mod. Phys. 84, 1527 (2012).

\bibitem{7}
I.Sick, Phys. Lett. B 576, 62 (2003).

\bibitem{8}
Blunden, P. G. and Sick, I. Phys. Rev. C 72, 057601 (2005).

\bibitem{9}
E. Kraus,1, K. E. Mesick, A. White, R. Gilman,1 and S. Strauch2, Phys. Rev. C90, 045206 (2014).

\bibitem{10}
Krzysztof M. Graczyk and Cezary Juszczak, Phys. Rev. C 90, 054334 (2014).

\bibitem{11}
Rong Wang, Xurong Chen, Phys. Rev. D 91, 054026 (2015).

\bibitem{12}
E. M Henley and G. A. Miller, Phys. Lett. B 251, 453 (1990).

\bibitem{13}
A.I. Signal, A.W. Schreiber and A.W. Thomas, Mod. Phys. Lett. A 6, 271 (1991);

\bibitem{14}
W. Melnitchouk, J. Speth, and A. W. Thomas, Phys. Rev. D59, 014033 (1998).

\bibitem{15}
N. N. Nikolaev, W. Schafer, A. Szczurek, and J. Speth, Phys. Rev. D 60, 014004 (1999).

\bibitem{16}
B.S. Zou, Nucl. Phys. A835, 199 (2010).

\bibitem{17}
B.S. Zou, D.O. Riska, Phys. Rev. Lett. 95 (2005) 072001.

\bibitem{18}
D.O. Riska, B.S. Zou, Phys. Lett. B 636 (2006) 265.

\bibitem{19}
C.S. An, D.O. Riska, B.S. Zou, Phys. Rev. C 73 (2006) 035207.

\bibitem{20}
S.J. Brodsky, P. Hoyer, C. Peterson, and N. Sakai, Phys. Lett. B 93, 451 (1980).

\bibitem{21}
Wen-Chen Chang and Jen-Chieh Peng, Phys. Rev. Lett. 106, 252002 (2011).

\bibitem{22}
M. Salajegheh, Phys. Rev. D 73, 074033 (2015).

\bibitem{23}
Keh-Fei Liu and Shao-Jing Dong, Phys. Rev. Lett. 72, 1790 (1994).

\bibitem{24}
Keh-Fei Liu, Phys. Rev. D 62, 074501 (2000).

\bibitem{25}
Keh-Fei Liu, Wen-Chen Chang, Hai-Yang Cheng, and Jen-Chieh Peng, Phys. Rev. Lett. 109, 252002 (2012).

\bibitem{26}
H.-L. Lai, M. Guzzi, J. Huston, Z. Li, P. M. Nadolsky, J. Pumplin, and C.-P. Yuan, Phys. Rev. D 82, 074024 (2010).

\bibitem{27}
A. D. Martin, W. J. Stirling, R. S. Thorne, and G. Watt, Eur. Phys. J. C 63, 189 (2009).

\bibitem{28}
Rong Wang, Xurong Chen, Chinese Physics C Vol. 41, No. 5 053103 (2017).

\bibitem{29}
G. Parisi and R. Petronzio, Phys. Lett. B 62, 331 (1976).

\bibitem{30}
A. I. Vainshtein, V. I. Zakharov, V. A. Novikov, and M. A. Shifman, JETP Lett. 24, 341 (1976).

\bibitem{31}
M. Glck and E. Reya, Nucl. Phys. B 130, 76 (1977).

\bibitem{32}
X. Chen, J. Ruan, R. Wang, P. Zhang, and W. Zhu, Int. J. Mod. Phys. E 23, 1450057 (2014).

\bibitem{33}
J. Pumplin, D. R. Stump, J. Huston, H.-L. Lai, P. Nadolsky, and W.-K. Tung, J. High Energy Phys. 07 012 (2002).

\bibitem{34}
Wei Zhu, Rong Wang, Jinghong Ruan et al. Eur. Phys. J. Plus 131, 6 (2016).

\bibitem{35}
K. A. Olive et al. (Particle Data Group), Chin. Phys. C, {\bf 38}, 090001 (2014).

\bibitem{36}
Yong-jun Zhang, Bin Zhang, Bo-Qiang Ma, Phys.Lett. B523 260-264 (2001).

\bibitem{37}
Yong-Jun Zhang, Wei-Zhen Deng, Bo-Qiang Ma, Phys.Rev. D65 114005 (2002) .

\bibitem{38}
J. Magnin and H. R. Christiansen, Phys. Rev. D61, 054006 (2000).

\bibitem{39}
Yun-hua Zhang, Lijing Shao, Bo-Qiang Ma, Phys.Lett. B671 30-35  (2009).

\bibitem{40}
E.Mac, E.Ugaz, Z. Phys.C 43 655 (1989).


\end{thebibliography}
\end{document}